\documentclass[letterpaper,journal]{IEEEtran}
\usepackage{amsmath,amsfonts}
\usepackage{algorithmic}
\usepackage{algorithm}
\usepackage{array}
\usepackage[caption=false,font=normalsize,labelfont=sf,textfont=sf]{subfig}
\usepackage{textcomp}
\usepackage{stfloats}
\usepackage{url}
\usepackage{verbatim}
\usepackage{graphicx}
\usepackage{cite}
\usepackage[hidelinks]{hyperref}
\usepackage{flushend}

\usepackage{booktabs}
\usepackage{mdframed}
\usepackage{xcolor}
\usepackage{multirow}
\usepackage{tabularx}
\usepackage{algorithm}
\usepackage{algorithmic}

\usepackage{threeparttable}

\usepackage{array}
\usepackage{multirow}
  
\usepackage{booktabs}
\usepackage{listings}
\usepackage{amsmath, bm}
\usepackage{color}

\usepackage{hyperref}

\begin{document}
%
\title{SolPhishHunter: Towards Detecting and Understanding Phishing on Solana}
%
%
%

\author{Ziwei Li, Zigui Jiang,~\IEEEmembership{Member,~IEEE,} Ming Fang, Jiaxin Chen, Zhiying Wu,\\Jiajing Wu,~\IEEEmembership{Senior Member,~IEEE,} Lun Zhang, Zibin Zheng,~\IEEEmembership{Fellow,~IEEE}

\thanks{The research is supported by National Natural Science Foundation of China under project (62472457), 
Guangdong Basic and Applied Basic Research Foundation (2023A1515011336). \textit{(Corresponding author: Zigui Jiang.)}} 

\thanks{Ziwei Li, Zigui Jiang, Ming Fang, Jiaxin Chen, Jiajing Wu and Zibin Zheng are with the School of Software Engineering, Sun Yat-sen University, Zhuhai 519082, China. Zhiying Wu are with the School of Computer Science and Engineering, Sun Yat-sen University, Guangzhou, 510006, China. Lun Zhang is with GoPlus Security.}

}

\maketitle

\begin{abstract}
Solana is a rapidly evolving blockchain platform that has attracted an increasing number of users. However, this growth has also drawn the attention of malicious actors, with some phishers extending their reach into the Solana ecosystem. Unlike platforms such as Ethereum, Solana has distinct designs of accounts and transactions, leading to the emergence of new types of phishing transactions that we term SolPhish. We define three types of SolPhish and develop a detection tool called SolPhishHunter. Utilizing SolPhishHunter, we detect a total of 8,058 instances of SolPhish and conduct an empirical analysis of these detected cases. Our analysis explores the distribution and impact of SolPhish, the characteristics of the phishers, and the relationships among phishing gangs. Particularly, the detected SolPhish transactions have resulted in nearly \$1.1 million in losses for victims. We report our detection results to the community and construct SolPhishDataset, the \emph{first} Solana phishing-related dataset in academia.
\end{abstract}


\section{Introduction}
Solana~\cite{Solana_whitepaper} is a high-performance blockchain platform renowned for its rapid transaction processing capabilities. It is capable of handling thousands of transactions per second, which has enabled Solana to secure a prominent position in the public blockchain sector. In 2024, Solana made a remarkable comeback from the depths of the 2022 crypto winter to become an industry leader. The price of Sol, the Solana native token, has gained 86\% in 2024 while soaring to a new all-time high of \$263.84 on Nov. 23. This price increase is confirmation that Solana crushed it last year. The total trading volume of the decentralized exchanges on Solana rose to an impressive \$626 billion in 2024, nearly catching up to Ethereum, which recorded \$674 billion~\cite{Solana_2024}.

As the popularity of Solana continues to soar, security concerns within the Solana ecosystem have garnered increasing attention. Regrettably, similar to other blockchains such as Ethereum, Solana is also susceptible to phishing scams. Phishing has long been a critical security issue in the blockchain domain. Scammers exploit users due to their unfamiliarity with blockchain technology or their momentary lapses in vigilance to illegally obtain private keys, thereby gaining control of user wallets or inducing them to sign malicious transactions. This results in significant financial losses for the victims. According to the annual report~\cite{certik_report} released by a security firm called Certik, phishing was the most damaging attack method on Web3 in 2024. Phishing scams have also emerged on Solana. As of January 2024, two major Solana wallet phishing tools—Rainbow Drainer and Node Drainer—had defrauded 3,947 victims of assets worth over \$4.17 million~\cite{drainer_report}. Similar phishing scams continue to occur on Solana. 



Although a number of studies~\cite{bilstm4dps,dynethnet,ewdps,novelhybrid,nft,payload,phishing_gangs,Siege,TTAGN,txphishscope,whoarephishers} have analyzed and detected phisher scams, existing research focuses primarily on phishing scams on Ethereum, achieving analysis and detection of Ethereum phishing accounts, transactions, and Web pages. To the best of our knowledge, no studies have yet investigated the phishing scams on Solana.

\begin{figure}[t]
  \centering
  \includegraphics[width=0.5\textwidth]{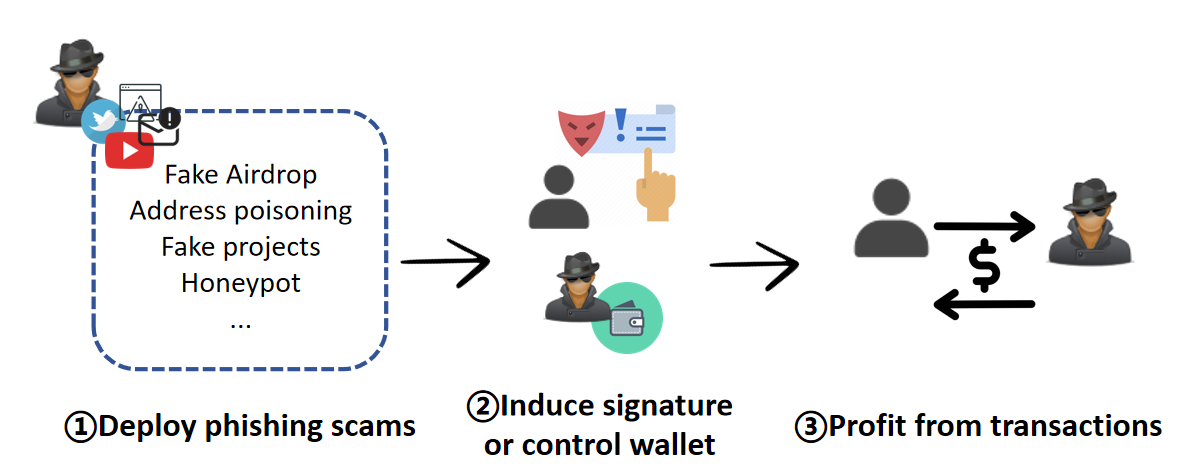}
  \caption{Solana phishing scam process}
  \label{phish}
\end{figure}

As shown in Fig.~\ref{phish}, phishing scams on Solana mainly consist of three key steps: \textbf{(1) Deploying phishing scams}: First, phishers disseminate phishing links or addresses through multiple channels, such as Twitter, YouTube, and email, to create and deploy phishing scams, including fake airdrops, address poisoning, fraudulent projects, and honeypots. \textbf{(2) Inducing signatures or controlling wallets}: Then, phishers induce victims to sign and confirm phishing transactions or deceive them into revealing their private keys to directly control their wallets. \textbf{(3) Profiting from transactions}: Finally, phishers profit by conducting malicious phishing transactions. It is worth noting that focusing on the "Profiting from transactions" step, and detecting phishing scams from the transaction level, is necessary. On the one hand, existing anti-phishing tools find it difficult to detect and defend against all phishing attacks in the first two steps. On the other hand, transaction profit is the most direct profit-making behavior and core objective of phishers, and it is also one of the key pieces of evidence of their illegal activities.

The process of phishing scams on Solana is similar to that on other blockchains. However, compared with other blockchains, Solana has different designs in transactions and accounts, which have given rise to some new types of phishing transactions on Solana. Specifically, Solana has the following characteristics:


\begin{itemize}
    \item \textbf{Transaction Design}: A notable feature of Solana transactions is the ability to include multiple instructions within a single transaction. Each instruction can complete an interaction. Therefore, a single Solana transaction with multiple instructions can facilitate multiple transfers. In contrast, on the EVM (Ethereum Virtual Machine), the transfer of each type of token requires a separate transaction.

    \item \textbf{Account Design}: Solana incorporates various types of accounts, including wallet accounts, token accounts, and program accounts. Each account type is controlled by a corresponding owner who manages permissions and operations. Additionally, Solana uses specific prefixes or suffixes (e.g., ``1111'') to distinguish between system official accounts (such as the system program and BPF loader) and other accounts. System official accounts are typically used to store system-level data or programs, such as the system program and token program.
\end{itemize}

Phishers have developed new phishing tactics, presented in Section \ref{sec:solphish}, based on the unique characteristics of Solana. We refer to these phishing transactions exploiting the design on Solana as \textit{\textbf{SolPhish}}. Due to the distinct transaction and account designed by Solana, the phishing transaction detection method~\cite{payload} previously used on Ethereum cannot be applied directly to detect SolPhish.

\textbf{Our Work.} Our research primarily focuses on analyzing and detecting phishing transactions that exploit the unique features of Solana, namely \textit{\textbf{SolPhish}}, which corresponds to the third step in the phishing process depicted in Fig.~\ref{phish}. We identify and summarize the new types of phishing transactions on Solana and propose \textit{\textbf{SolPhishHunter}}, a detection tool based on effective patterns for identifying SolPhish. We then conduct empirical analysis based on detected phishing transactions. We delineate the scope of our work from two dimensions and differentiate our work from existing research: \textbf{(1) Platform}: We concentrate on phishing scams on Solana. To the best of our knowledge, we are the \emph{first} to analyze and detect Solana phishing transactions. \textbf{(2) Granularity}: We primarily focus on the novel types of phishing transactions employed by scammers on Solana, rather than on how scammers induce victims to sign transactions. Our detection is transaction-level, targeting the new types of phishing transactions on Solana. Phishing transaction types that already exist on other platforms, such as Ethereum, are not the primary focus of our work.

\textbf{Contributions.} In summary, the contributions of our work are as follows:

\begin{itemize}
    \item To the best of our knowledge, we are the \emph{first} to investigate phishing transactions on Solana and have summarized three unique types of phishing transactions, i.e., Single Transaction with Multiple Transfers, Account Authority Transfer, and Impersonation of System Account.

    \item We propose SolPhishHunter, which employs effective detection rules for the three identified types of SolPhish, achieving an overall detection precision of 93.96\%.

    \item Using SolPhishHunter, we detect 8,058 phishing transactions and construct the first Solana phishing transaction dataset, SolPhishDataset, providing a foundational data resource for research in this area.
    
    \item Based on the detected SolPhish transactions, we conduct an empirical analysis to examine their temporal distribution, financial losses incurred, characteristics of phishing accounts, and potential collusion among attackers, followed by a summary of several key findings.

    \item We report a total of 64 phishing accounts and 8058 phishing transactions to Solscan and the security firm Goplus, contributing to anti-phishing efforts within the Solana ecosystem.
    
\end{itemize}


The remainder of this paper is organized as follows. Section \ref{sec:Prelim} presents some preliminaries related to this study. Section \ref{sec:solphish} defines the three types of SolPhish identified. Section \ref{sec:SolPhishHunter} presents the proposed detection tool SolPhishHunter and demonstrates the detection results with further evaluation on SolPhishHunter. Section \ref{sec:charaterizing} introduces the empirical analysis based on the detected SolPhish. Section \ref{sec:contribution} discusses our contributions to the community. Section \ref{sec:discussion} provides a discussion of our work and Section \ref{sec:related} reviews the related works. Finally, the paper is concluded in Section \ref{sec:conclustion}.
\section{Preliminary}\label{sec:Prelim}

\subsection{Solana Accounts}



\textbf{(1) Wallet Accounts.}
Wallet accounts (also known as user accounts) store native token on Solana, SOL, which is used to pay for transaction fees and account rent. These accounts serve as the foundational means for users to interact with the Solana network, similar to wallet addresses on other blockchains. Wallet accounts are created and managed by the system program, and the holder of the private key associated with the wallet account has permissions, including asset control, authorization of operations, and transaction signing.

\textbf{(2) Token Accounts.}
Token accounts are used to store and manage specific tokens. In the Solana ecosystem, in addition to the native SOL token, there are numerous other tokens issued on the Solana blockchain, such as governance tokens for various DeFi projects and stablecoins. Each token has its corresponding token account, and users must create the appropriate token accounts within their wallets to hold and manage these tokens. Token owners have permissions to transfer or burn the tokens.

\textbf{(3) Program Accounts.}
Program accounts are used to store and execute smart contract program code on Solana. The smart contracts on Solana, known as ``Programs,'' are stored in program accounts in bytecode form. Program accounts can receive and process transaction instructions from other accounts and execute the corresponding smart contract logic. For example, they can implement various functions of decentralized applications (DApps), such as creating and managing digital assets, executing trading logic, and processing user interactions.

\subsection{Solana Transactions}


Users interact with the network or other users by sending transactions. Solana transactions consist of several key data items:

\textbf{(1) Instructions.}
Instructions are the smallest operational units within a transaction, with different instruction types corresponding to different operations. Common instructions include transfer, assign, createAccount, and setAuthority. A single transaction can contain multiple instructions, which are executed sequentially but can be processed in parallel through the Sealevel runtime (by analyzing account dependencies between instructions and processing non-conflicting instructions in parallel).

\textbf{(2) Recent Block Hash.}
When creating a transaction, users must obtain the latest block hash from a validator node and submit the transaction promptly to avoid expiration. During transaction execution, the hash is verified to ensure it is within the recent block set (typically valid for 150 blocks, approximately one minute). If the hash is outdated, the transaction will be rejected. The recent block hash prevents transaction replay attacks and ensures the timeliness of transactions.

\textbf{(3) Signers.}
Each transaction must be signed by at least one private key holder to authenticate the legitimacy of the operation (e.g., a transfer requires the signature of the sender). In scenarios requiring multi-party authorization, such as organizational accounts or governance operations, some transactions need multiple signers to jointly authorize the transaction, known as multi-signature transactions. Signers are also responsible for paying transaction fees, which are correlated with the consumption of computational resources.

\section{Categorizing SolPhish}\label{sec:solphish}

\begin{figure}[t]
  \centering
  \includegraphics[width=0.48\textwidth]{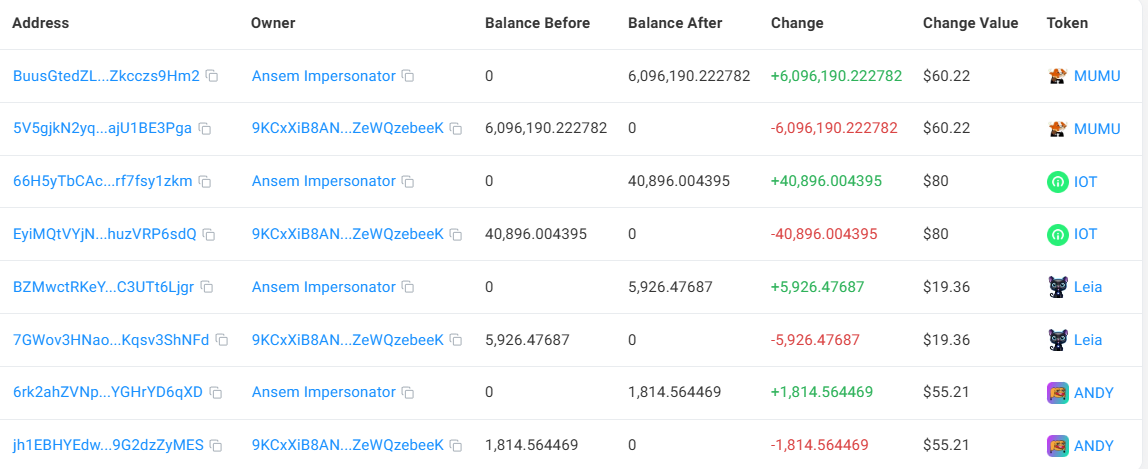}
  \caption{The phishing transaction f2MA...PaiC succeeded in draining multiple types of tokens from the wallet of victims.}
  \label{STMT_example}
\end{figure}

From the perspective of luring methods, phishing scams on Solana predominantly employ tactics such as fake airdrops, counterfeit project websites, free lotteries, promises of high returns, and address poisoning to induce users to sign phishing transactions or to gain control of their wallets. These deceptive methods are constantly evolving, yet they bear similarities to the common phishing lures observed on EVM-based platforms. However, from the transaction level, the unique design of transactions and accounts on Solana has provided scammers with new means of conducting their schemes. We have identified three types of phishing transactions based on the characteristics of Solana (i.e., SolPhish), which specifically include single transaction with multiple transfers, account authority transfer, and impersonation of system accounts.

\subsection{\textbf{Single Transaction with Multiple Transfers, STMT}}
A single transaction on Solana can encompass multiple instructions, each of which can be utilized to achieve distinct functions (such as transferring tokens or creating accounts). Leveraging this characteristic, phishers are not required to set up separate transfer transactions for each type of token, as they would on Ethereum. Instead, they can plunder the wallet of a victim by including multiple transfer instructions within a single malicious transaction. Fig.~\ref{STMT_example} illustrates the changes in token balances for both the victim and the phisher in a phishing transaction (f2MA...PaiC). This transaction comprises five \textit{transfer} instructions, which facilitate the plundering of four types of tokens and SOL coins from the wallet of victims.

\subsection{\textbf{Account Authority Transfer, AAT}}

\begin{figure}[t]
  \centering
  \includegraphics[width=0.5\textwidth]{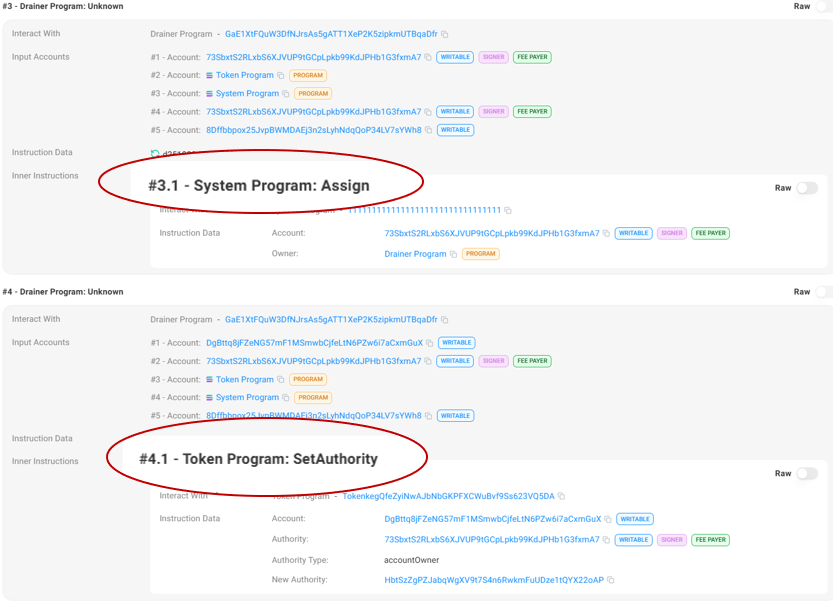}
  \caption{SolPhish based on AAT (4GVr...tYAn)}
  \label{aat_example}
\end{figure}

Solana encompasses a variety of account types, including wallet accounts and token accounts, each with a corresponding owner who controls the permissions of this account. Scammers can induce victims or gain control of their wallets to sign authority transfer transactions, thereby taking possession of the assets of victims. Specifically, for wallet accounts, phishers can use transactions with the \textit{Assign} instruction to set the owner of a wallet account to a phishing program they have meticulously deployed. For token accounts, phishers can employ the \textit{SetAuthority} instruction to set themselves as the owner of the targeted token account. Note that these two types of authority transfers may occur within the same transaction. For instance, Fig.~\ref{aat_example} illustrates a phishing transaction (4GVr...tYAn) including two authority transfer instructions. Specifically, first, the phisher uses the \textit{Assign} instruction to set the owner of the targeted wallet account to a meticulously deployed phishing program. Then, by interacting with this phishing program and using \textit{setAuthority} instruction, the phisher sets themselves as the owner of the targeted token account. Finally, the phisher deploys another fund transfer transaction\footnote{Fund transfer transaction: 43MVswMUJwvjPAZqTPbKmGXaCbqshfQCPnBre9rEaEpmh3CbooYaivVw9k9cyVLSVnztVuBbQhySmQfVXA3AFDqy} to move the tokens of the victim to multiple different phishing accounts.


\subsection{\textbf{Impersonation of System Accounts, ISA}}


On Solana, the official system accounts are distinguished from other accounts through unique prefixes and suffixes. However, some malicious scammers exploit this feature by employing the `solana-keygen grind' command or the Solana tool SlerfTools\footnote{https://slerf.tools/zh-cn/vanity-address-generator/solana} to generate vanity addresses that resemble system accounts (i.e., addresses containing specific prefixes and suffixes). These scammers then induce victims to sign transactions that interact with these counterfeit addresses. TABLE~\ref{Phishng_accounts} illustrates some of these spurious system account addresses. Given that these addresses contain similar prefixes and suffixes (e.g., ``11111'' and ``Compu''), and most wallet software only displays partial prefixes and suffixes of the interacting addresses, victims may fail to carefully examine the addresses and mistakenly assume that the interaction is with an official account, thus trusting the transaction.

\begin{table}[t]
\centering
\caption{Some system accounts and phishing accounts}
\begin{tabularx}{0.45\textwidth}{lX}
\toprule

\begin{tabular}[c]{@{}l@{}}
     System\\Accounts
\end{tabular}
&
\begin{tabular}[c]{@{}l@{}}
111111111111111111111111111111111 \\
ComputeBudget111111111111111111111111111111 \\
NativeLoader111111111111111111111111111111  
\end{tabular}
\\
\midrule

\begin{tabular}[c]{@{}l@{}}
     Phishing\\Accounts
\end{tabular}
& 
\begin{tabular}[c]{@{}l@{}}
Qcsb89L6QS74b56...CnBSqnQ5gq11111 \\
CaNCU6LiZUKc7Su...eAmv625c4M11111 \\
TdnNjtovxBmRZmg...a4CCfTUNxU11111 \\
np84cd63UoFj2pb...w6j2hN4poa11111 \\
iBGtY2LBEmTiVrm...EmmkDxbLhV11111 \\
CompuV3LmCTW7AG...ZKkK1fCAY8L7eM1 
\end{tabular}
\\
\bottomrule

\end{tabularx}
\label{Phishng_accounts}
\end{table}
\section{Detecting SolPhish}\label{sec:SolPhishHunter}

To detect SolPhish, based on the analysis presented in the preceding section, we formulate specific detection rules for each type of phishing transaction and propose a SolPhish detection tool, SolPhishHunter.

\subsection{Design of SolPhishHunter}

\begin{table*}[t]
\centering
\caption{Rules of SolPhishHunter}
\begin{tabular}{c|l}
\toprule
\textbf{Types} & \textbf{Rules} \\ 
\midrule
Prerequisites & 
\begin{tabular}[c]{@{}l@{}}
tx.from $\notin$ Markets \\
$\land$ tx.to $\notin$ Markets \\
$\land$ $\neg$ CONTAINS(tx.log, ``buy" $\lor$ ``sell" $\lor$ ``purchase") \\
$\land$ tx.from == tx.to
\end{tabular} \\ \midrule
STMT & 
\begin{tabular}[c]{@{}l@{}}
(COUNT(TransferIns $\in$ tx.inss) $>$ 2) \\
$\land$ (COUNT(\{ tb $\in$ tx.tokenbalance $|$ tb.balance\_pre!=0 $\land$ tb.balance\_after == 0 \}) $\geq$ 2)
\end{tabular} \\ \midrule
AAT & 
\begin{tabular}[c]{@{}l@{}}
CONTAINS(tx.inss, AssignIns) \\
$\lor$ (CONTAINS(tx.inss, SetAuthorityIns) \\
$\land$ SetAuthorityIns.authorityType == ``account owner")
\end{tabular} \\ \midrule
ISA & 
\begin{tabular}[c]{@{}l@{}}
CONTAINS(tx.inss, TransferIns) \\
$\land$ ($\exists$ tb $\in$ tx.tokenbalance $|$ tb.balance\_pre!=0 $\land$ tb.balance\_after == 0 \\
$\lor$ ($\exists$ sb $\in$ tx.solbalance $|$ sb.balance\_pre!=0 $\land$ sb.balance\_after == 0)) \\
$\land$ (tx.to.address MATCHES (``Compu.*'' $\lor$ ``.*1111''))
\end{tabular} \\ \bottomrule
\end{tabular}
\label{rules}
\end{table*}

The rules we establish are shown in TABLE~\ref{rules}. In the table, CONTAINS(a, b) denotes that set or string a includes element b. ``Markets'' refers to trading markets. TransferIns, AssignIns, and SetAuthorityIns correspond to instructions of the types Transfer, Assign, and SetAuthority, respectively. tx.inss, tx.log, tx.tokenbalance, tx.solbalance refer to the instruction set of the transaction, the transaction log, the records of changes in token account balances and the records of changes in Sol tokens involved in the transaction, respectively. tx.from and tx.to indicate the loser and the beneficiary of the transaction, respectively. In the phishing scenarios involving STMT and ISA, tx.from refers to the account that suffers a loss of funds, while tx.to refers to the account that gains the funds. In the phishing scenario involving account authority transfer, tx.from refers to the original owner of the account authority, and tx.to refers to the new owner of the account authority.

\subsubsection{Prerequisties}
For transactions under scrutiny, we first eliminate normal market activities, including token exchanges, NFT purchases, and sales. These transactions may involve instructions related to phishing activities (e.g., authority transfers) but do not fall within the scope of our defined SolPhish. Therefore, we establish four preliminary rules to avoid false positives for these benign transactions. A transaction is deemed non-phishing if it meets any of the following four conditions: (1) The beneficiary of the transaction is a trading market; (2) The loser of the transaction is a trading market; (3) The transaction log contains keywords related to market activities, such as ``buy,'' ``sell," or ``purchase"; (4) The beneficiary and the loser of the transaction are the same, which typically indicates an exchange of funds or authority.

\subsubsection{Rules for STMT}

To detect phishing transactions involving multiple transfers within a single transaction, we set rules in two aspects: (1) Multiple Transfers: The transaction instruction set should contain more than two \textit{transfer} instructions; (2) Wallet Draining: The transaction involves the depletion of two or more types of tokens. This rule is primarily used to reduce the likelihood that the transaction is a routine user transfer, as users rarely transfer all their tokens of multiple types in normal circumstances.

\subsubsection{Rules for AAT}

To detect phishing transactions involving account authority transfers, we add rules related to instruction usage based on the preliminary rules. For wallet account authority transfers, we check if the transaction contains an Assign instruction. For token account authority transfers, we check if the transaction contains a SetAuthority instruction with the authorityType parameter set to ``account owner." Note that the strict preliminary rules have already filtered out transactions using these instructions for benign market activities, such as transferring tokens or NFTs via SetAuthority.

\subsubsection{Rules for ISA}

For counterfeit system account phishing, we set rules in two main aspects: (1) Fund Transfer: The transaction involves the depletion of SOL tokens or a specific token from the victim's account; (2) Suspicious Address: The beneficiary's address contains the prefix ``Compu" or the suffix ``11111," which raises suspicion of impersonating an official system account. The former increases the likelihood of phishing by involving token depletion, while the latter helps determine if the beneficiary is suspected of impersonating an official account.

\subsubsection{Implementation}
\begin{figure*}[t]
  \centering
  \includegraphics[width=0.9\textwidth]{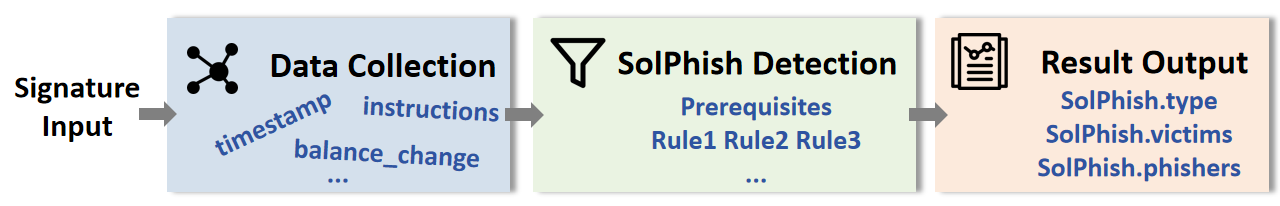}
  \caption{The framework of SolPhishHunter}
  \label{SolPhishHunter}
\end{figure*}

We implement the prototype of SolPhishHunter using Python. As shown in Fig.~\ref{SolPhishHunter}, SolPhishHunter consists of three key modules. \textbf{(1) Data Collection Module}: For transactions under scrutiny, we use the Solana data crawling interface. Drawing on the RPC-based ETL tool proposed by Wu et al.~\cite{knowyourtransactions}, we employ request-level parallel calls to enhance the speed of data acquisition. \textbf{(2) SolPhish Detection Module}: We implement the aforementioned rules to detect and output results based on the crawled transaction information. \textbf{(3) Result Output Module}: For suspicious transactions, we output the SolPhish type, the victims, and the phishers.

\subsection{Evaluation of SolPhishHunter}\label{sec:evaluation}

In this subsection, we will explore the effectiveness of SolPhishHunter.

\subsubsection{\textbf{Datasets}} The dataset used for evaluation primarily consists of the historical transactions of accounts. Initially, we collect datasets of phishing accounts and normal accounts. Subsequently, we crawl the historical transactions of these accounts to form the transaction datasets for phishing accounts and normal accounts, respectively.


\textit{\textbf{(i) Phishing Account Dataset.}}
We collect phishing accounts from two sources:

\begin{itemize}
    \item \textit{Chainabuse~\cite{chainabuse}}: Chainabuse is a platform that allows users to submit their victimization reports and share their experiences with phishing scams. We collect 166 victim reports under the Solana phishing tag from this platform in January 2025. We manually review each report, filtering out those with insufficient details (which make it impossible to determine the presence of phishing) and those that do not provide phishing addresses. This process yields a total of 138 reports containing 148 phishing accounts.

    \item \textit{Solscan~\cite{solscan}}: Solscan is a Solana data browser which labels phishing accounts based on complaints from social media and reports from the security community. These labels serve as warnings to Solana users to reduce the risk of financial loss. We search for addresses containing keywords such as ``fake\_phishing" and ``drainer" on Solscan, collecting a total of 27 phishing accounts.

\end{itemize}

After removing duplicates (6 accounts) from the accounts collected through these two sources, we construct a phishing account dataset comprising 169 unique addresses.

\textit{\textbf{(ii) Normal Account Dataset.}} 
We construct the normal account dataset by selecting the top 200 accounts with the highest asset values from the Account Leaderboard on Solscan. These accounts are chosen for the following reasons: 
\begin{itemize}
    \item \textit{Diverse Account Types.} The top 200 accounts include both mainstream project accounts and private accounts, representing a wide range of account types.
    
    \item \textit{Diverse Transaction Types.} These accounts typically engage in a variety of benign transactions, including market trades, mutual transfers, and token exchanges, reflecting a broad spectrum of normal transaction behaviors.

    \item \textit{Low Phishing Probability.} These accounts are often key opinion leaders (KOLs) whose transactions are influential and serve as references for other users. Their transactions are less likely to be associated with phishing activities.
\end{itemize}

\textit{\textbf{(iii) Transaction Dataset.}}
Based on the phishing account dataset and the normal account dataset, we crawl the historical transactions of these accounts. This process results in the creation of two datasets:  the \textbf{Transaction Dataset of Phishing Accounts (TDPA)}, containing 262,719 transactions, and the \textbf{Transaction Dataset of Normal Accounts (TDNA)}, containing 274,705 transactions.

The information of the datasets used in the evaluation is summarized in TABLE~\ref{datasets}.

\begin{table}[t]
\centering
\caption{The information of datasets}
\begin{tabular}{c|cc}
\toprule
\textbf{Dataset}&\textbf{Number of Accounts}&\textbf{Number of Transactions}\\
\midrule
\textbf{TDPA}&169&262,719 \\
\textbf{TDNA}&200&274,705 \\
\bottomrule
\end{tabular}
\label{datasets}
\end{table}

\subsubsection{\textbf{Experimental Setup}}
To evaluate the effectiveness of the established rules, we conduct rule-based detection on TDPA and TDNA. We assess the effectiveness of SolPhishHunter from two dimensions: (i) whether SolPhishHunter can detect phishing transactions from phishing accounts; (ii) whether SolPhishHunter generates false positives for benign transactions from normal accounts.

\subsubsection{\textbf{Evaluation Results}}

We apply SolPhishHunter to detect phishing transactions in the two transaction datasets mentioned earlier. For each transaction under test, the experimental results are shown in TABLE~\ref{detect_results}.

\begin{table}[t]
\centering
\caption{Detection results of SolPhishHunter}
\begin{tabular}{ccc}
\toprule
\textbf{Type}&\textbf{TDPA(TP / Detected)}&\textbf{TDNA(TP / Detected)}\\

\midrule

\textbf{STMT}&2438 / 2773&0 / 0 \\
\textbf{AAT}&2171 / 2272&0 / 5 \\
\textbf{ISA}&3449 / 3526&0 / 0 \\

\textbf{Total}&\textbf{8058 / 8571}& \textbf{0 / 5} \\
\bottomrule
\end{tabular}
\label{detect_results}
\end{table}

\textbf{Results on TDFA}:
In the transaction dataset of phishing accounts, a total of 8,571 suspicious transactions are detected, accounting for approximately 3.26\%. Among these detected suspicious transactions, 2,774 (32.35\%) involve single transactions with multiple transfers, 2,272 (26.51\%) involve account authority transfers, and 3,526 (41.14\%) involve impersonation of system accounts.

Note that 3.26\% represents the proportion of SolPhish detected among all historical transactions of the labeled phishing accounts. While this number is not particularly high, it is reasonable for the following reasons: (1) Even for phishing accounts, the majority of transactions are not phishing transactions but rather normal fund transfers; (2) This study does not aim to detect all phishing transactions conducted by phishers. Instead, it focuses on identifying three novel types of phishing transactions that arise from the unique design of Solana, namely STMT, AAT and ISA. However, some phishing transactions that are not included in the detection scope also exist on other blockchains and have been studied and analyzed. These methods~\cite{payload, adresspoisoning} (although based on other platforms) may be extendable to Solana. 
(3) Some phishing transactions may not be distinguishable from benign transfers at the transaction level and thus fall outside the scope of our detection.

Based on the types of beneficiaries and losers in the transactions, these 8,573 suspicious transactions are labeled and classified. Of these, 8,058 transactions (94.01\%) are identified as phishing transactions, 130 transactions represent mutual transfers between labeled phishing accounts, and 383 transactions involve transfers of funds or authority from labeled phishing accounts to other entities (suspected money laundering transactions).

\textit{\textbf{(i) Transactions based on STMT}}

Among the 2,773 detected transactions, 2,438 are identified as phishing transactions, and 80 represent mutual transfers between labeled phishers. 28 phishing transactions involve the use of labeled phishing programs to drain multiple types of tokens from victims. A total of 25 phishing accounts have conducted phishing scams through single transactions with multiple transfers. The most prolific among them is Gck5...1VX4\footnote{Gck5PWhKL4Qn87bhFwpL19Y5gkAbFN93GXXsTzuJ1VX4}, with 1,639 phishing transactions associated with it. This account has also been reported by multiple Twitter users~\cite{Twitter_gck1}\cite{Twitter_gck2} for malicious wallet-draining activities.

In the remaining 245 suspicious transactions, funds flow from 12 labeled phishing accounts to other accounts, which we consider as money laundering transactions. The most prominent money laundering accounts are GLqf...3bUh\footnote{GLqfgNJmVYBeT9TrEpvZPf7t79Jd8EFrsCti87RA3bUh} and HbtS...2oAP\footnote{HbtSzZgPZJabqWgXV9t7S4n6RwkmFuUDze1tQYX22oAP}. The former transfers funds to 16 different addresses through 124 single transactions with multiple transfers, while the latter transfers funds to another account through 89 such transactions. We find relevant tweets~\cite{Twitter_glqf}\cite{Twitter_hbts} about these two phishing accounts on Twitter, which also point out that they drain victims' wallets and subsequently transfer the funds.

Additionally, an interesting finding is that 14 of the detected single transactions with multiple transfers use the \textit{advanceNonce} instruction. This is related to the durable nonce in the Solana ecosystem. Durable nonce is a mechanism to bypass the short recent\_blockhash lifespan of transactions to prevent transaction expiration. Users can deploy and send offline transactions on Solana using durable nonce, meaning these transactions can be executed at a later time. For phishers, this allows them to deploy phishing transactions that do not execute immediately, so the victims' funds are not lost right away, making the phishing activity less likely to be detected. Such phishing behavior deserves careful attention and scrutiny.

\textbf{\textit{(ii) Transactions based on AAT}} 

Among the 2,272 detected suspicious transactions involving account authority transfers, 2,171 are phishing transactions, all of which include wallet account authority transfers. Of these, 1,391 transactions involve both wallet account authority transfers and token account authority transfers.

For transactions involving wallet account authority transfers, the beneficiaries of all transactions point to three labeled phishing programs, which helps us confirm that these transactions are indeed profit transactions of phishing accounts. Among them, the phishing program BNRT...5Rep\footnote{BNRThTYg9x49JYNkbDUYERb3X7JV2GYcEpFLAUHX5Rep} has the largest volume of related phishing transactions, with 931 transactions involving the transfer of account authority.

Among the transactions involving token account authority transfers, the beneficiaries of 1,391 transactions point to labeled phishers. The remaining 101 transactions involving token account authority transfers involve transferring token account authority from labeled phishers to other accounts. Upon further examination of these 101 transactions, we find that the beneficiaries of these funds have frequent financial dealings with phishing accounts. Therefore, we consider these beneficiaries to be part of the same gang as the labeled phishing accounts. In these 101 transactions, we identify a total of five gang accounts.

\textbf{\textit{(iii) Transactions based on ISA}} 

Among the 3,526 detected suspicious transactions involving impersonation of system accounts, 3,449 are phishing transactions with beneficiaries pointing to eight labeled phishers. Fifty transactions represent mutual transfers between labeled phishers. 

The remaining 27 transactions involve the transfer of funds from labeled phishing accounts to other unlabeled accounts. However, these accounts receiving the funds also feature the same prefixes or suffixes as official system accounts. Therefore, despite not being labeled as phishing accounts, we consider them to be part of the same gang as the labeled impersonating accounts. Specifically, we identified 13 distinct phishing accounts from these 27 transactions. We list these impersonating accounts in the TABLE~\ref{fake_accounts}.

\begin{table*}[h]
    \centering
    \caption{Labeled phishing accounts and discovered phishing gangs}
    \begin{tabular}{c|c}
    \toprule
    \textbf{Labeled Phishing Accounts} & \textbf{Discovered Phishing Gangs}\\
    \midrule
        \begin{tabular}[c]{@{}l@{}}
            iBGtY2LBEmTiVrmPCgHRGdCPZJcDEmmkDxbLhV11111\\
            gYs5v8LBaTNRFhU8rSSFeEqaCQPjmkT28naAN711111\\
            CompuV3LmCTW7AGGnM6YBftCJkKP3ZKkK1fCAY8L7eM1\\
            np84cd63UoFj2pbR7QU226hKg8e5w6j2hN4poa11111\\
            CaNCU6LiZUKc7Su3avu5jDbdDXdKeAmv625c4M11111\\
            Qcsb89L6QS74b56PpxvygajsXKx1CnBSqnQ5gq11111\\
            TdnNjtovxBmRZmg76oHhTgpVfWQYa4CCfTUNxU11111\\
            b8pyTBjatpNCVaKxRbPkqiujziAcvTXkcQeYDb11111\\
        \end{tabular}
    &   \begin{tabular}[c]{@{}l@{}}
            9LMa3PfmGCA61itb479zZVU9Pjqg8j1fXzP1Ex11111\\
            eWxJCpJh1ac5gv3j7i6GM6FvhndkZP1j1bebNc11111\\
            cMHG3ivVkVmkUcX6MMsWetEUeWcCw78QzSaqzW11111\\
            Lm4Ng7FPwxGBVQAPYNbqijuUzcXiy9vaVWPAPE11111\\
            ZJKJzZpiEXkK2CSGi6eJYGvZ6bp3kRXwPBsiWV11111\\
            CompuMqjzwwtYH5kFbFgqFW92UJzNRSUXMBud1G4svh2\\
            CompurwaRQEp7NAbdSnHhmu35gNm3HwGRdBtVSVFjcX5\\
            wddBT1UoieJbHt3gHrtEWEyh41jDe4PfctgpGs11111\\
            ComputHPT5daVapYegd9NEuqWfHdy9DAcG2Yiskc8uX8\\
            CompuN1ccmbcd2wfDU8j8bkLG62AGgHAwMx4XRnGWKTM\\
            RB8EsMj3dntJZtotspEfPisei5oQjTB4f4NWjx11111\\
            S8dSx4oehDpy7CVexQ2HVMepdoxnZF6puEbdwH11111\\
            De2bQHxEgbiQCp4T8E1vrNXy8x19tMs49yWtVu11111\\
            
        \end{tabular}\\
    \bottomrule
    \end{tabular}
    \label{fake_accounts}
\end{table*}

\textbf{Results on TDNA}: 
On TDNA, SolPhishHunter flags five transactions as phishing transactions involving account authority transfers. These five transactions are further analyzed by three independent blockchain experts who assess whether they are phishing transactions. All experts conclude that none of the five transactions are phishing transactions, prompting a false positive analysis.

Four of these five transactions involve using the \textit{Assign} instruction to set the owner of a wallet account to Rout...xxpb\footnote{RouterBmuRBkPUbgEDMtdvTZ75GBdSREZR5uGUxxxpb}. Upon reviewing the transaction context on Solscan and the code repository~\cite{Jito-tip-router} of this program, it is identified as an MEV Tip Router program related to Jito. Jito is a liquid staking service on Solana that helps users submit transactions more efficiently and at a lower cost through intelligent transaction ordering and bundling strategies. This program is used to handle the distribution of transaction fees. However, since the program is not labeled by Solscan as a project or exchange program (lacking a public name) and matches the rule of transferring wallet account authority to an unknown program, SolPhishHunter flags it as a phishing transaction.

The remaining transaction involves a user claiming PYTH tokens. The core operations include verifying user eligibility, creating an associated token account, and transferring tokens from a pool to the account of user. In this transaction, a temporary storage account, Euio...Lu7c\footnote{EuioRpEkSeYhamHfbNAZ6RTpsPx72xh4ro8VBx8JLu7c}, receives SOL from 3Lrh...YXJu and then transfers ownership to the EXxq...e9ch program. It is speculated that this account is used solely to temporarily hold funds and ensure that the account is used strictly according to specific program logic through the `Assign` instruction, with subsequent automatic allocation of funds by the program. However, the lack of subsequent transactions for this wallet account limits further analysis.


\textbf{Summary:} SolPhishHunter marks a total of 8,576 suspicious transactions, among which 8,058 are SolPhish transactions, with a precision of approximately \textbf{93.96\%}. Overall, on the one hand, SolPhishHunter can detect phishing transactions in suspicious phishing accounts; on the other hand, the historical transactions of normal accounts are basically not judged as phishing transactions. These prove the rationality and effectiveness of the rules we have formulated.

\section{Charaterizing  SolPhish}\label{sec:charaterizing}


In this section, we delve deeper into the phishing transactions detected by TDFA to uncover the characteristics and evolving trends of Solana phishing scams. Specifically, we investigate the temporal distribution of SolPhish, the economic losses SolPhish cause, the characteristics of the phishers, and the relationships within phishing gangs.

\subsection{\textbf{Temporal distribution patterns}}

\begin{figure}[t]
  \centering
  \includegraphics[width=0.5\textwidth]{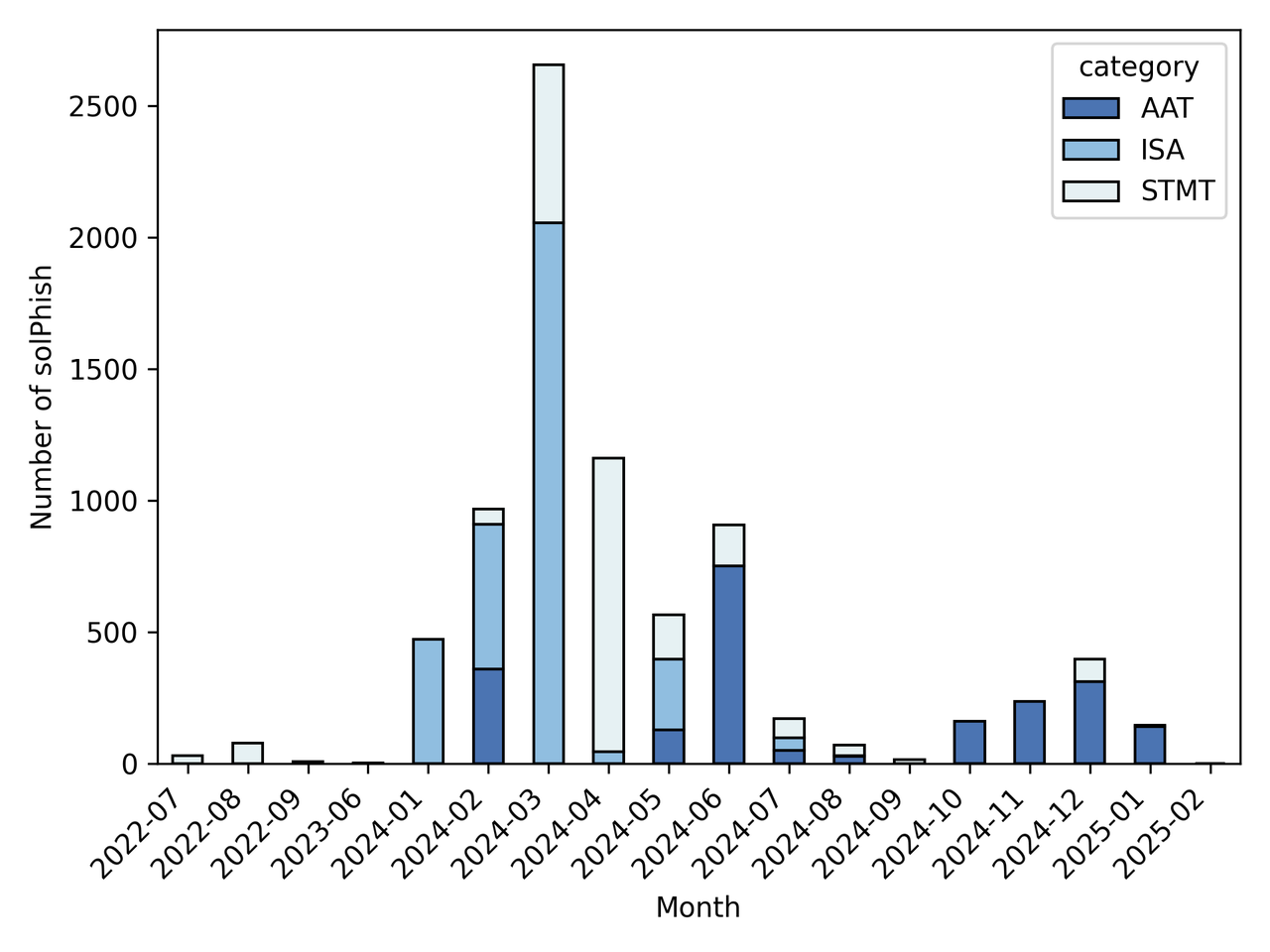}
  \caption{Time distribution of SolPhish}
  \label{time}
\end{figure}

To examine the temporal distribution of detected phishing transactions, we analyze the number of transactions detected each month based on the time of occurrence. The number of detected phishing transactions for each type across different months is shown in Fig.~\ref{time}. In terms of the time distribution of SolPhish, we find that:

\textbf{(1) Overall temporal distribution.} 
The volume of phishing transactions peaks from January to June 2024. The majority of detected SolPhish transactions concentrate in this six-month period (accounting for 83.62\%), with the highest number (2,657) occurring in March 2024. This may correlate with rapid development of Solana. The year 2024 marks a market recovery for Solana, characterized by a significant increase in user activity and an overall upward trend in the value of its native token SOL in the first half of the year. The positive development attracts not only more users to the Solana ecosystem but also the attention of greedy phishers. However, since some phishers are exposed and multiple security companies publish warnings about phishing scams~\cite{solphish_report1,solphish_report2,solphish_report3}, the phishing situation eases somewhat in the second half of the year.

\textbf{(2) Evolution of phishing types.} 
The three types of SolPhish exhibit a certain evolutionary trend, generally progressing from ISA to STMT to AAT. This is likely because, as users' security awareness increases and wallet security measures continue to improve, phishers are forced to evolve their phishing methods. Their approaches are shifting towards more concealed and technically sophisticated means.

Impersonation of System Accounts (ISA) emerges first in large numbers. This type of phishing transaction relies on having the same prefixes and suffixes as official system accounts, with a relatively low implementation cost. The design of some wallets that only display the prefixes and suffixes of transfer addresses also facilitates this type of phishing. However, these phishing accounts have obvious characteristics and are easy to detect and discover, so they become less common after June 2024.

Single Transaction with Multiple Transfers (STMT) follows. This method usually requires more up-front inducement to lure users into signing transactions that will drain their wallets, and it is more common in March-April 2024. This is closely related to the emergence of Solana drainer, a malicious tool or script specifically used to steal cryptocurrency wallet assets. Attackers use this tool to transfer all tokens, NFTs, and other assets from the target wallet to addresses they control in a very short time by inducing users to sign malicious transactions or authorizations. The core purpose of the attackers is to ``drain" the assets from the wallet of users. Single Transaction with Multiple Transfers (STMT) is one of the important ways for them to achieve this. Since January, there have been several reports~\cite{drainer_report}\cite{drainer_report1} about the emergence and malicious behavior of Solana drainer.

Account Authority Transfer (AAT) appears relatively later. In most cases, the permissions of both the wallet account and the token account are transferred by the phishers at the same time. Unlike Ethereum, where token authorization is usually required first before further fund transfers, and both authorization and fund transfers need to be signed by the token owner, the \textit{Assign} and \textit{SetAuthority} instructions in Solana directly change the account owner. Phishers gain control of the account directly, and the consequences of users signing transactions containing authority transfer instructions are more direct and severe. This type of phishing method is more likely to deceive new Solana users, especially those who do not understand the Solana permission mechanism.

\begin{mdframed}[linewidth=0.7pt, linecolor=black, skipabove=10pt, skipbelow=10pt, backgroundcolor=gray!6]
\textbf{Finding 1.} SolPhish is primarily concentrated in the first half of 2024, with phishing transactions exhibiting an evolving trend from ISA to ATMT to AAT.
\end{mdframed}

\subsection{\textbf{Financial loss}}


To investigate the financial losses caused by SolPhish to victims, we calculate the losses for each transaction. For convenience, we query the latest prices of various tokens using the token query interface provided by Oklink~\cite{oklink}. For phishing transactions based on STMT and ISA, we calculate the value of the tokens transferred by the victims. For AAT-based phishing transactions, although tokens in the account may not be transferred in the transaction, the phishers gain control of the corresponding token accounts due to authority transfer. Therefore, we calculate the total value of the tokens in the accounts with transferred authority as the victims' losses.

\begin{figure}[t]
  \centering
  \includegraphics[width=0.5\textwidth]{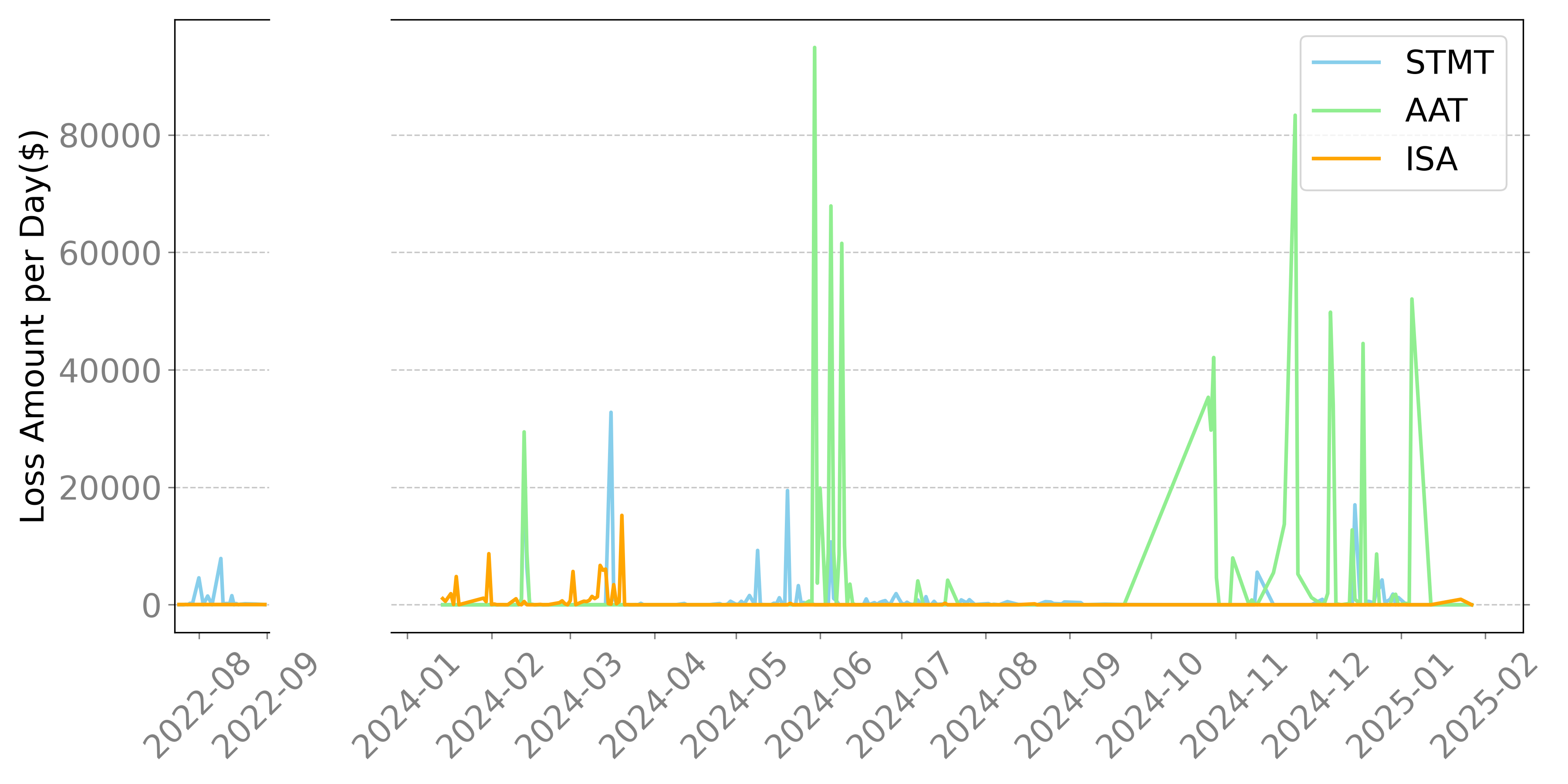}
  \caption{The losses caused to victims of SolPhish}
  \label{loss_fig}
\end{figure}

The daily financial losses caused by the detected SolPhish are presented in Figure~\ref{loss_fig}, with losses from different types of SolPhish represented by lines of different colors. A total of 8,058 detected phishing transactions result in nearly \$1.07 million (\$1,099,815.98) in losses for victims, with an average loss of \$136.49 per transaction. Furthermore, the pattern of financial losses aligns with the temporal distribution of phishing activities discussed in finding1, which exhibits a similar evolutionary trend (ISA-STMT-AAT).

\begin{table}[t]
    \centering
    \caption{Losses casued by each type of SolPhish}
    \begin{tabular}{llll}
        \toprule
        \textbf{Type} & 
        \textbf{Total Losses}& 
        \textbf{Average Loss} & 
        \textbf{Highest Loss}\\
        
        \midrule
        \textbf{STMT} & $\$211,894.26$ & $\$86.91$ & $\$150,19.12$\\

        \textbf{AAT} & $\$812,219.75$ & $\$374.12$ & $\$751,88.51$\\

        \textbf{ISA} & $\$757,01.98$ & $\$21.95$ & $\$106,34.25$\\
        \bottomrule
    \end{tabular}
    \label{loss_tabel}
\end{table}

The total financial losses caused by each type of phishing transaction, the average loss per transaction, and the maximum loss from a single transaction are shown in Table~\ref{loss_tabel}. Overall, \textbf{SolPhish based on AAT causes the most significant financial damage}, resulting in cumulative losses of \$812,219.75. This is consistent with expectations, as AAT directly transfers ownership and control of the account to the phishers, effectively causing the loss of all tokens under the account of victims. Although AAT phishing transactions represent the smallest proportion of all detected phishing transactions, they account for 73.85\% of the total financial losses, and this type of phishing has become more prevalent in recent times. In contrast, despite the higher number of ISA-related phishing transactions, the financial losses they cause are relatively lower (\$75,701.98).

\begin{table}[t]
    \centering
    \caption{Top 10 most targeted tokens}
    \begin{tabular}{cccc}
        \toprule
        \textbf{Top} & \textbf{Token} & \textbf{Price} & \textbf{Phish\_times} \\
        \midrule
        1 & \textbf{USDC} & 0.9998 & 332 \\
        2& \textbf{Jupiter} & 0.493113107 & 141 \\
        3&\textbf{Bonk} & 9.82353E - 06 & 139 \\
        4&\textbf{Wen} & 2.19188E - 05 & 123 \\
        5&\textbf{WIF} & 0.442559228 & 86 \\
        6&\textbf{Tooker Kurlson} & 0.00044323 & 86 \\
        7&\textbf{MOTHER IGGY} & 0.005599556 & 73 \\
        8&\textbf{USDT} & 0.99971 & 69 \\
        9&\textbf{BEER} & 4.03639E - 06 & 69 \\
        10&\textbf{Infinity} & 159.2746384 & 61 \\
        \bottomrule
    \end{tabular}
    \label{top10tokens}
\end{table}


Furthermore, we examine the primary tokens targeted by these phishing transactions. In addition to the native SOL token, these phishing transactions involve the transfer of 10,308 different types of tokens. For each token, we count the number of phishing transactions involving that token. The top 10 most targeted tokens are shown in TABLE~\ref{top10tokens}. Analysis of TABLE~\ref{top10tokens} reveals some interesting findings:

\textbf{Stablecoins as core phishing targets.} USDC (332 times) and USDT (69 times), as leading stablecoins, occupy two of the top 10 spots, with USDC at the top. This reflects that phishers tend to target assets that directly peg to fiat currency values, which have short monetization paths, stable prices, and excellent liquidity. Despite their low unit prices, stablecoins are held in large quantities by users, and a single attack may transfer assets worth thousands or even tens of thousands of dollars, offering a high risk-reward ratio. This finding is similar to the conclusion in Chainalysis’s 2025 report on cryptocurrency crime trends~\cite{2025_Crypto_Crime_Trends}, which states that stablecoins now account for the majority of all illicit transaction volumes (63\% of all illicit transactions). 


\textbf{Concentrated risks in Meme coins.} Meme coins are cryptocurrencies inspired by internet pop culture or social trends, whose core value relies more on community sentiment, social media propagation, and speculative trading rather than actual technological applications or economic models. Among the top 10 most targeted tokens in Table~\ref{top10tokens}, four are Meme coins, namely BONK, WEN, WIF, and BEER. This is likely because, despite their extremely low unit prices (all below \$0.01), these coins have a large number of holders and highly active communities, providing phishers with a broad scope for phishing activities.

\begin{mdframed}[linewidth=0.7pt, linecolor=black, skipabove=10pt, skipbelow=10pt, backgroundcolor=gray!6]
\textbf{Finding 2.} A total of 8,058 phishing transactions resulted in nearly 1.07 million in losses for victims, with SolPhish based on AAT causing the most economic damage. Stablecoins and meme coins have become the primary targets of phishing attacks.
\end{mdframed}

\subsection{\textbf{SolPhish phishers}}


To investigate the characteristics of phishers from the perspective of phishing accounts, we examine the number of phishing attempts, the losses caused, and the lifecycle of all phishing accounts. First, we count the number of phishing attempts for each account and the losses incurred by victims from these transactions. We show two Top 10 lists: one for the accounts with the highest number of phishing attempts (TABLE ~\ref{top10phishers_times}) and another for the accounts causing the most financial losses (TABLE ~\ref{top10phishers_loss}). We find that:

\begin{table}[t]
    \centering
    \caption{Top 10 phishers with the highest number of phishing attempts}
    \begin{tabular}{ccccc}
        \toprule
        \textbf{Top} & \textbf{Account} & \textbf{Times} & \textbf{Loss Amount} &\textbf{Phishing Type}\\
        \midrule
        1&CaNC...1111&1692&\$43547.24685&ISA\\
        2&Gck5...1VX4&1639&\$716.632956&STMT\\
        3&BNRT...5Rep&931&\$449934.8012&AAT\\
        4&GaE1...aDfr&881&\$300492.3516&AAT\\
        5&3Bit...7yQa&647&\$449636.1267&AAT\\
        6&Comp...7eM1&604&\$10412.64626&ISA\\
        7&HbtS...2oAP&596&\$300232.4992&AAT\\
        8&mRf8...1Nnk&359&\$39539.36887&AAT\\
        9&b8py...1111&336&\$4959.732427&ISA\\
        10&GLF8...epBJ&227&\$16430.8815&STMT\\
        \bottomrule
    \end{tabular}
    \label{top10phishers_times}
\end{table}

\begin{table}[t]
    \centering
    \caption{Top 10 phishers causing the most losses}
    \begin{tabular}{ccccc}
        \toprule
        \textbf{Top} & \textbf{Account} & \textbf{Times} & \textbf{Loss Amount} &\textbf{Phishing Type}\\
        \midrule
        1&BNRT...5Rep&931&\$449934.8012&AAT\\
        2&3Bit...7yQa&647&\$449636.1267&AAT\\
        3&GaE1...aDfr&881&\$300492.3516&AAT\\
        4&HbtS...2oAP&596&\$300232.4992&AAT\\
        5&6xWo...5XZK&220&\$50394.26012&STMT\\
        6&DmJa...gr4r&51&\$45506.66756&STMT\\
        7&CaNC...1111&1692&\$43547.24685&ISA\\
        8&mRf8...1Nnk&359&\$39539.36887&AAT\\
        9&Fwtf...SWYE&148&\$39538.7564&AAT\\
        10&JDrX...9v8z&57&\$32348.17048&STMT\\
        \bottomrule
    \end{tabular}
    \label{top10phishers_loss}
\end{table}

\textbf{(1) Phishing attacks based on STMT exhibit significant volatility in profitability.} Some accounts have a high number of phishing attempts but relatively low financial losses (e.g., Gck5...1VX4~\footnote{Gck5PWhKL4Qn87bhFwpL19Y5gkAbFN93GXXsTzuJ1VX4}). In contrast, some phishing accounts cause substantial financial losses with only a few phishing transactions (e.g., DmJa...gr4r~\footnote{DmJaFCGnvQK3txE5UpMRxe2i4yZ8FEfSGr3Tsk4Lgr4r}). In other words, STMT-based phishing can be either high-frequency and low-efficiency or low-frequency and high-efficiency. This is related to the phishing method itself, as STMT-based phishers primarily aim to drain wallets rather than target specific tokens in specific wallets. Therefore, the financial losses often correlate with the victims' wallet balances and the types of tokens held, leading to significant fluctuations in the losses per transaction caused.

\textbf{(2) Phishing attacks based on AAT show centralized and large-scale harm.} We find that among the top 10 accounts with the highest number of phishing attempts, half use AAT as their method. Similarly, among the top 10 accounts causing the greatest financial losses, six employ AAT. AAT-based phishing involves large-scale phishing accounts with both high frequency and significant financial impact, which warrants close attention. This aligns with our findings2 that AAT-based phishing results in substantial financial losses.

\begin{figure}[t]
  \centering
  \includegraphics[width=0.5\textwidth]{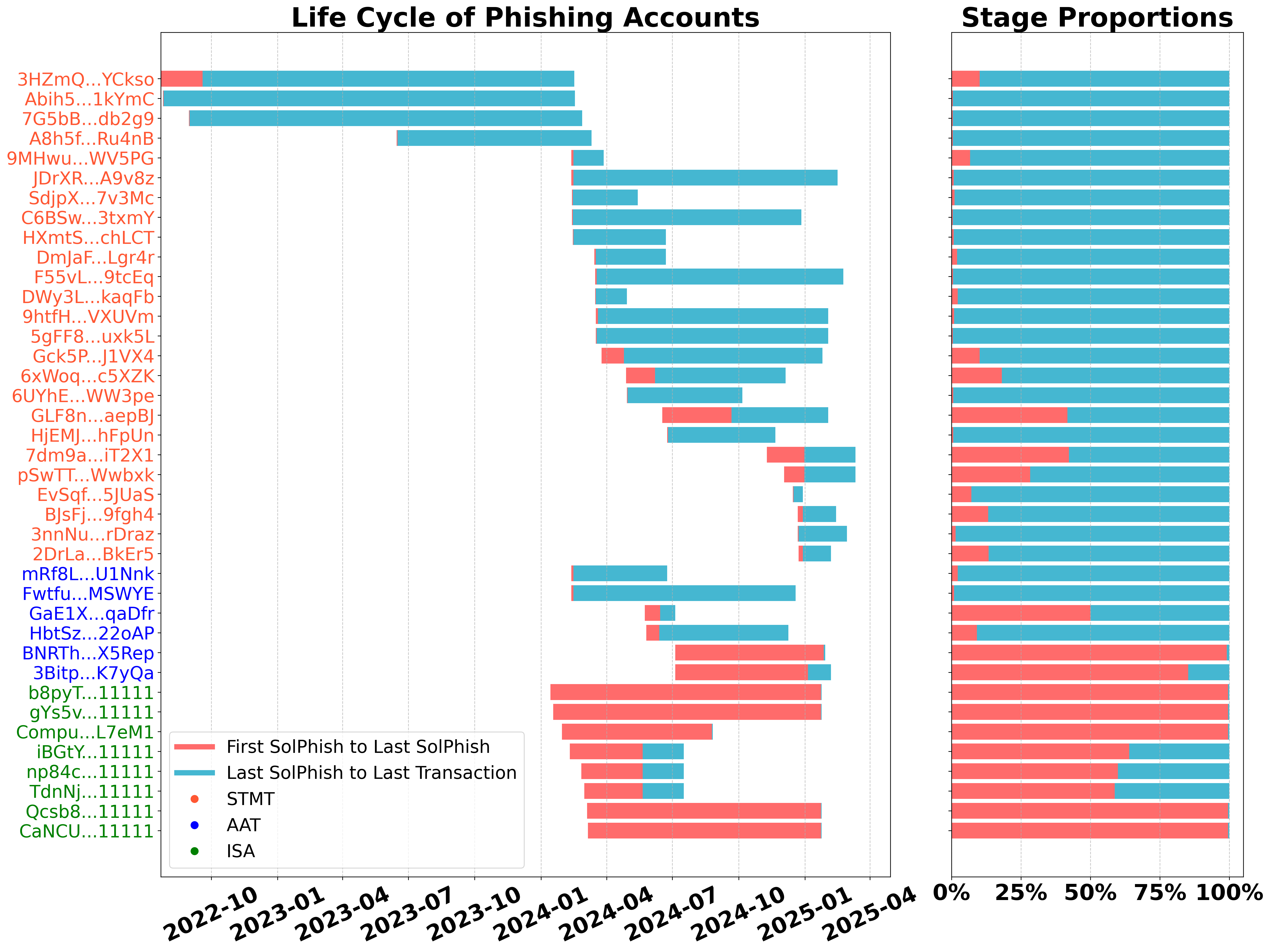}
  \caption{The life cycles of phishers}
  \label{life cycle}
\end{figure}


In addition, we further investigate the life cycles of these phishing accounts. For each phishing account, we record the time of the first detected phishing transaction, the time of the last detected phishing transaction, and the time of the most recent transaction made by the account. For convenience, we refer to the period from the first phishing transaction to the last phishing transaction as the phishing period, and the period from the last phishing transaction to the most recent transaction as the dormant period (note that ``dormant'' here means no phishing activity, not necessarily no transactions).

We present the phishing periods and dormant periods of various phishing accounts in Fig.~\ref{life cycle} (left subplot), as well as the proportion of these two periods within their overall life cycles (right subplot). Based on our statistical results, we draw the following conclusions:

\textbf{Overall Distribution}: Except for a small number of STMT-based phishing accounts, the majority of accounts have life cycles primarily distributed in 2024 and beyond. Among them, 82.5\% of the accounts have been inactive in the past month (after February 12, 2025). Moreover, accounts using the same phishing method tend to have similar life cycles and distributions.

\textbf{STMT-Based Phishing Accounts}: These accounts generally have shorter phishing periods, indicating that phishers cease phishing activities after a brief period. However, these accounts continue to engage in transactions for a relatively long time after their last phishing attempt.

\textbf{ISA-Based Phishing Accounts}: These accounts generally have longer phishing periods, with the longest extending up to approximately one year. ISA-based accounts quickly become inactive after their last phishing transaction, forming a contrasting pattern to STMT-based phishing accounts.

\begin{mdframed}[linewidth=0.7pt, linecolor=black, skipabove=10pt, skipbelow=10pt, backgroundcolor=gray!6]
\textbf{Finding 3.} The profits of phishers based on STMT are unstable, while the damage caused by phishers based on AAT is characterized by being concentrated and large-scale. The vast majority of phishing accounts become inactive within a month, with phishers based on STMT having a shorter phishing cycle.
\end{mdframed}

\subsection{\textbf{Phisher gangs}}


To investigate whether these phishing accounts are associated with syndicates, we scan the historical transactions of the labeled phishing accounts and detect any interactions among them. We consider two types of interactions: the first is direct fund transfer, which involves transferring tokens via the \textit{transfer} instruction; the second is authority transfer, which involves transferring tokens via the \textit{setAuthority} instruction.

\begin{figure}[t]
  \centering
  \includegraphics[width=0.36\textwidth]{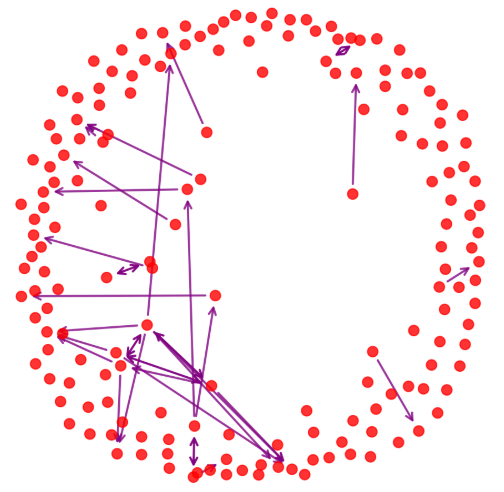}
  \caption{Relationship betweent the labeled phishers}
  \label{relation}
\end{figure}

\begin{figure}[t]
  \centering
  \includegraphics[width=0.5\textwidth]{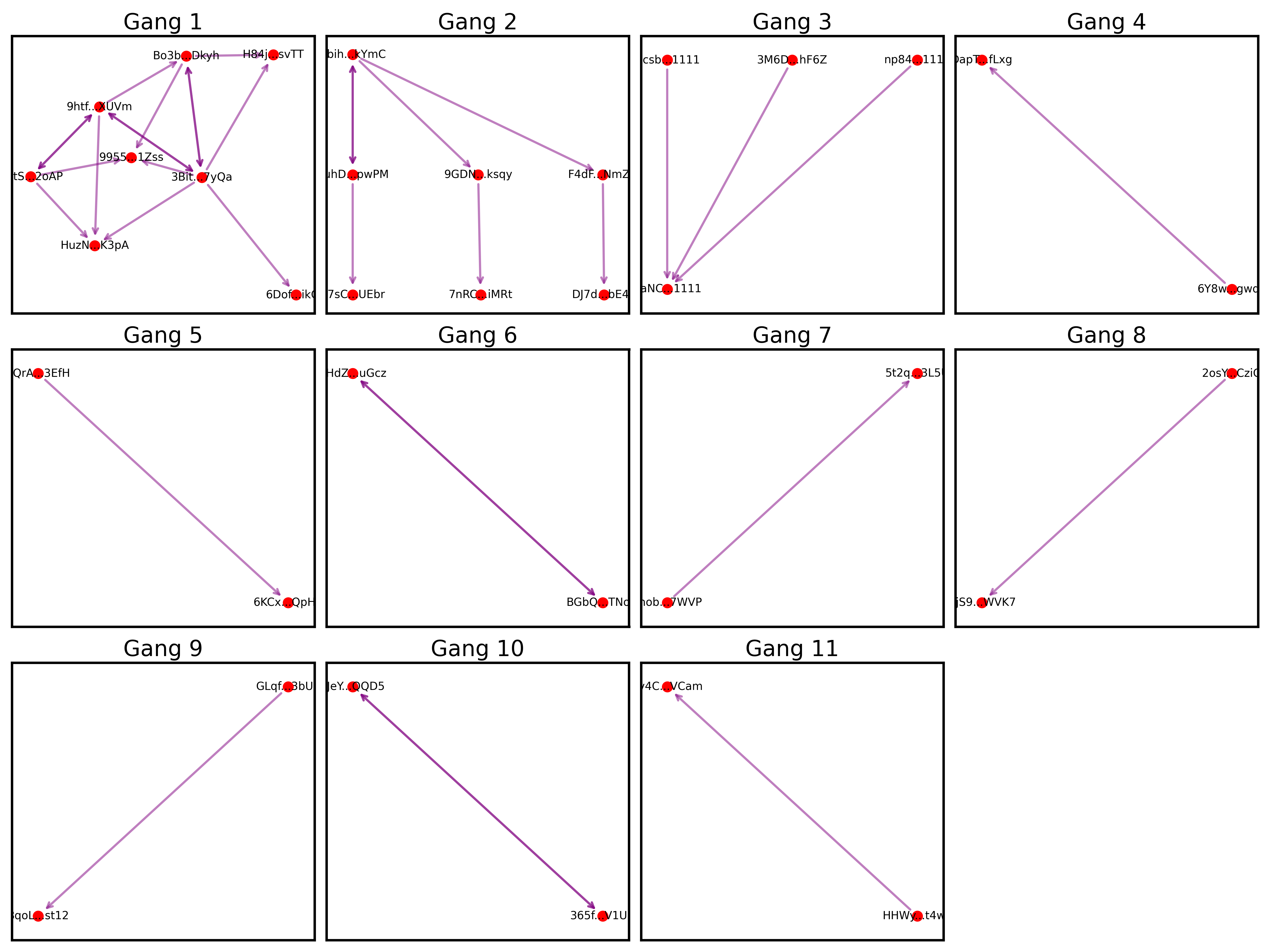}
  \caption{Phisher gangs}
  \label{phisher gangs}
\end{figure}

We construct a network, as shown in Fig.~\ref{relation}, depicting the interactions among these labeled phishing accounts. In this network, edges represent historical interactions between accounts, and arrows indicate the direction of the interactions. It is evident that multiple accounts have interaction histories, forming suspicious phishing gangs. Further analysis of Fig.~\ref{relation} reveals 11 distinct gangs. We then conduct an in-depth analysis of the three largest gangs.

\begin{table}[t]
    \centering
    \caption{Members of Gang1, Gang2 and Gang3}
    \begin{tabular}{c|c}
        \toprule
         Gang & Phishers \\
         \midrule
         Gang1 & \begin{tabular}[c]{@{}c@{}}
\textbf{9htfHryds3YLLjZZxa4RTyegFfsScvGcFTQyFCcVXUVm}\\ \textbf{3Bitp9awjSEGE1UumthfF6iogTuTkFhaLBHj3xvK7yQa}\\ \textbf{HbtSzZgPZJabqWgXV9t7S4n6RwkmFuUDze1tQYX22oAP}\\ 6Dofca9F1pU2iodvgzGFSUaP8BSooMsWPb8H5saYikCX\\ H84jE7ZvSQc7JM6BpZX5kBKDShYboM6yYLaAZKdGsvTT\\  Bo3bz5HiGB8V9vxDPc8p6CujtXdosT1c4CvU3wdYDkyh\\
9955r2AsxJcbuST11jeE39wGXeQX5FziNhjzeJMK1Zss\\ HuzNeiRA6MexzQs21AKHUsjLWTseCNkTcFi5iX9MK3pA\\
         \end{tabular}
         \\
         \midrule
         Gang2 & \begin{tabular}[c]{@{}l@{}}
\textbf{Abih5VzqFR8qMCJZNi1C7pJjkVZThgaKk9r6gjv1kYmC}\\
6uhD4cWsiFgRaH7KC39waEVb6quJprgrUhkCTH65pwPM\\ 67sCNziG6MPms4RfpPofVjRB31gMM2YMbYXRYRbYUEbr\\ DJ7doi4vLtL8N3kpQjh1YK91gcK1a4JYmeUPtJL4bE4K\\ 7nRCFwUnZpPYbP4T9hsAqaUaqu2RHx9VhBuvfLoLiMRt\\ 9GDN5krysAR8xGnLaRZTvzcPzfmForf1VjJzUqCSksqy\\ F4dF9bnf5WxcfTJnDxt4D4t2CipseQgESXS24YCnNmZ\\
         \end{tabular}
         \\
         \midrule
         Gang3 & \begin{tabular}[c]{@{}l@{}}
\textbf{Qcsb89L6QS74b56PpxvygajsXKx1CnBSqnQ5gq11111}\\
\textbf{np84cd63UoFj2pbR7QU226hKg8e5w6j2hN4poa11111}\\
\textbf{CaNCU6LiZUKc7Su3avu5jDbdDXdKeAmv625c4M11111}\\ 3M6DXwQva1H3oTSiYUJCjPSFqTYzQWm3rMiPpQjzhF6Z\\
         \end{tabular}
         \\
         \bottomrule
    \end{tabular}
    \label{tab:my_label}
\end{table}


\textbf{Gang 1} comprises eight labeled phishing accounts. Within the historical transactions of three of these accounts (9htf...XUVm, etc.), a total of 1,246 SolPhish transactions were detected, resulting in losses of \$749,879.60 for the victims. The primary phishing methods employed by this gang include STMT and AAT. Within the gang, a distinct flow of funds originating from the account 3Bit...7yQa and transferring to other phishing accounts has been identified, forming an overall star-shaped network topology with a radiating structure.


\textbf{Gang 2} comprises seven labeled phishing accounts. Within the historical transactions of one of these accounts (Abih...kYmC), a total of 21 SolPhish transactions were detected, resulting in losses of \$51.22 for the victims. The primary phishing method employed by this gang is account permission transfer. Within the gang, a distinct flow of funds originating from the account Abih...kYmC and transferring to other phishing accounts has been identified, forming an overall hierarchical tree-like topology.

\textbf{Gang 3} comprises four labeled phishing accounts. Within the historical transactions of three of these accounts (Qcsb...1111, etc.), a total of 2,071 SolPhish transactions were detected, resulting in losses of \$466,279.20 for the victims. The primary phishing method employed by this gang is impersonation of system accounts. Unlike other gangs, Gang 3 has formed a unified flow of funds towards the account CaNC...1111, creating a pattern of separate phishing activities followed by centralized fund aggregation. Most members of this gang impersonate system accounts, with “1111” as the account suffix. Overall, the gang forms an inwardly radiating star-shaped topology.


\begin{mdframed}[linewidth=0.7pt, linecolor=black, skipabove=10pt, skipbelow=10pt, backgroundcolor=gray!6]
\textbf{Finding 4.} Some gang relationships exist among the labeled phishers, with a total of 11 gangs identified. The three largest gangs involve 19 phishing accounts in total.
\end{mdframed}

\section{Contribution to the Community}\label{sec:contribution}


To enhance the security of the Solana community, this paper explore the following directions:

\subsection{Data Disclosure}

\begin{figure}
    \centering
    \includegraphics[width=1.0\linewidth]{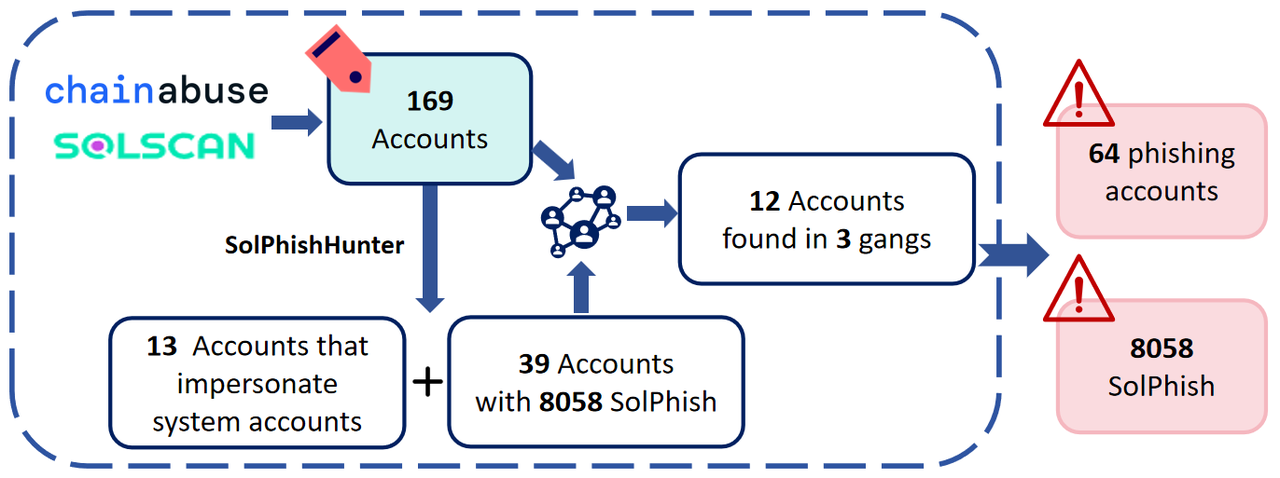}
    \caption{Composition of SolPhishDatset}
    \label{sol_datasets}
\end{figure}

We collect the detected phishing accounts, their associated phishing transactions, and the relationships among these accounts, and publicly release \textit{\textbf{SolPhishDataset}}, the \textit{\textbf{first}} dataset in academia related to Solana phishing scams. This dataset comprises 64 phishing accounts and 8,058 phishing transactions, which we hope will provide a basis for future research on Solana security. Fig.~\ref{sol_datasets} reviews how SolPhishDataset is constructed. First, we collect 169 accounts that have been reported or flagged by Solscan as phishing accounts. Then, using SolPhishHunter, we have detected 39 accounts with 8,058 SolPhish transactions in their historical transaction set TDPA. Additionally, we identify 13 accounts that are also impersonating system accounts but have not been flagged (as shown in TABLE~\ref{fake_accounts}). Furthermore, in our exploration of RQ4 in Section VI, we uncover connections among the 169 accounts collected and identify 12 accounts within three relatively larger groups that have transaction histories with the accounts where SolPhish is detected. Although SolPhish is not detected in the historical transactions of these 12 accounts, they have been reported by users and have interacted with the detected SolPhish accounts. Therefore, we consider them part of the phishing gangs. In total, we obtain 64 phishing accounts and 8,058 phishing transactions.

\subsection{Reporting to the Community}


 Among the 64 phishing accounts, 46 accounts (71.88\%) have not been flagged by Solscan~\cite{solscan}. We report these accounts to Solscan, hoping that Solscan’s flagging of these accounts will help more users understand their nature, reduce the risk of being phished, and avoid potential losses. In addition, we collaborate with the security firm Goplus, providing them with the phishing tactics and detection results that we have uncovered. Goplus has acknowledged our findings and published a relevant report to alert the Solana user community to the phishing scams detailed in this paper.

\subsection{Recommendations}


To help more Solana users reduce the risk of being phished, we propose the following recommendations for future applications.

\begin{itemize}
    \item \textbf{Enhance Security Awareness}: Some phishing tactics exploit the negligence and lack of understanding of the Solana ecosystem, especially among new users. Therefore, before engaging with the Solana ecosystem, users should thoroughly understand Solana’s design and the potential attack vectors. By raising their security awareness and remaining vigilant, users can avoid clicking on unknown links or buttons.

    \item \textbf{Verify Interaction Addresses}: The best time to conduct due diligence on the counterparty before signing any transaction is six months ago; the second-best time is today. Users should carefully verify the counterparty before signing a transaction. We recommend checking whether the counterparty has been reported or flagged as a risky account. We also call on wallet software or security plugins to provide users with risk perception functions that can alert users in a timely manner when they are about to interact with risky accounts, thereby helping users avoid risks.

    \item \textbf{Utilize Transaction Simulation Features}: Some wallets offer transaction simulation features, which users can utilize to carefully review the simulation results and check for any unreasonable transfers or authorizations.
    
\end{itemize}

\section{Discussion}\label{sec:discussion}

\subsection{Limitation}


This study currently focuses on three types of phishing transactions, which may not cover all possible types of Solana phishing. In terms of scope, we primarily concentrate on novel phishing transaction types arising from Solana’s unique characteristics, namely the definition and detection of three types of SolPhish. However, our detection tool, SolPhishHunter, is designed to be easily extensible. When new types of phishing transactions are identified, corresponding detection rules can be readily added to enable the detection of these emerging phishing transactions.

\subsection{Future Work}

\textbf{Integration with Wallets}: One of our future work is to integrate SolPhishHunter with wallets. When a user is about to sign a transaction, our tool will monitor and analyze the transaction to assess the risk of phishing. It will then provide timely feedback to the user, helping them perceive potential risks and thereby reducing economic losses.

\textbf{Exploring Additional Phishing Types}: Phishing tactics are constantly evolving, and attackers continually develop new methods. A key area of our future work will be to investigate novel phishing techniques on the Solana platform and, where possible, propose detection solutions to further combat fraud within the Solana ecosystem.

\subsection{Ethical Considerations}


We emphasize that our research does not involve any privacy concerns. First, the data employed in our research are sourced exclusively from the publicly accessible blockchain transaction records on Solana, which are available to anyone. Second, the dataset we release consists solely of anonymized data, with no inclusion of personally identifiable information. Given these considerations, we are assured that our research does not compromise any privacy policy or result in any form of legal action against individuals.

\section{Related Work}\label{sec:related}

\subsection{Phishing Scams Detection}




Phishing scams are one of the most common frauds on blockchain platforms and have caused significant economic losses. Numerous existing studies have been dedicated to the detection of phishing scams.

In the detection of phishing accounts, Wu et al.~\cite{whoarephishers} pioneeringly proposed the trans2vec network embedding algorithm, which employs a random walk-based method to extract and classify features of Ethereum phishing accounts, marking the first step in the detection of Ethereum phishing accounts. Subsequently, a series of studies~\cite{TTAGN, Siege, novelhybrid} model the transactions between accounts on Ethereum as a graph and utilize neural networks to integrate node attributes and transaction network topology features, thereby achieving phishing account detection. More recent works~\cite{dynethnet, bilstm4dps, ewdps} further focus on the dynamic evolution of the transaction network, incorporating contextual information of transaction records between accounts to improve the accuracy of phishing account detection.

In addition to phishing account detection, other studies investigate phishing gangs, NFT phishing scams, phishing websites, and phishing transactions on Ethereum. Liu et al.~\cite{phishing_gangs} conducted the first study to formalize the problem of phishing gang detection, using high-risk phishing accounts as a starting point and employing genetic algorithm optimization to uncover potential phishing gangs. Yang et al.~\cite{nft} systematically analyzed NFT phishing scams for the first time, summarized the methods used in NFT phishing, and conducted empirical analysis based on collected cases. He et al.~\cite{txphishscope} focused on phishing web pages on Ethereum and proposed the first Ethereum phishing web page detection system, TxPhishScope. Chen et al.~\cite{payload} analyzed payload-based transaction phishing on Ethereum and proposed a rule-based phishing transaction detection method.

Our work distinguishes itself from the aforementioned research in the following aspects: (1) \textit{Different platform.} We focus on phishing scams on Solana for the first time. (2)\textit{Different Granularity.} We mainly investigate three novel types of phishing transactions based on the unique characteristics of Solana.

\subsection{Solana Security}



Solana has experienced rapid development in recent years. Concurrently, some researchers have also begun to focus on security issues within the Solana ecosystem. Cui et al.~\cite{vrust} implemented a static program analysis-based framework for detecting novel vulnerabilities in Solana smart contracts. Smolka et al.~\cite{fuzz} proposed the first fuzzing framework for Solana smart contracts based solely on binaries, enhancing the accuracy and usability of vulnerability detection in Solana smart contracts. In addition, Andreina et al.~\cite{odds} summarized the security challenges faced by Solana developers and the actual vulnerabilities in the Solana ecosystem, and validated the important role of automated verification of the Anchor framework in ensuring smart contract security.

In contrast to the above work, our focus is primarily on transaction security rather than contract security on Solana.

\section{Conclusion}\label{sec:conclustion}

This work is the first to explore phishing scams on Solana, analyzing the unique characteristics of account and transaction design on Solana and identifying three types of phishing transactions based on these characteristics, namely SolPhish. Subsequently, we proposed SolPhishHunter, which effectively detected 8,058 instances of SolPhish in phishing accounts. We then conducted an empirical analysis of the detected phishing transactions, exploring the distribution of SolPhish, the economic losses caused, the characteristics of the phishers, and the relationships among phishing gangs. Finally, we constructed SolPhishDataset, the first Solana phishing-related dataset in academia, and reported our detection and analysis results to the community. In the future, we will continue uncovering phishing scams within the Solana ecosystem, improve our phishing detection tools, and contribute to the security of the Solana ecosystem.

%
\IEEEpeerreviewmaketitle
\bibliographystyle{IEEEtran}
\bibliography{bibfile}{}




\end{document}